# Spin-resolved quantum-dot resonance fluorescence

A. Nick Vamivakas[1,†], Yong Zhao[1,2,†], Chao-Yang Lu[1,3], & Mete Atatüre[1]

[1]*Cavendish Laboratory, University of Cambridge, JJ Thomson Avenue, Cambridge CB3 0HE, UK*

[2]*Physikalisches Institut, Ruprecht-Karls-Universität Heidelberg, Philosophenweg 12, 69120 Heidelberg, Germany*

[3]*HFNL & Department of Modern Physics, University of Science and Technology of China, Hefei, 230026, China*

[†]*These authors contributed equally to this work.*

**In the quest for physically realizable quantum information science (QIS) primitives, self-assembled quantum dots (QDs) serve a dual role as sources of photonic (flying) qubits and traps for electron spin; the prototypical stationary qubit. Here we demonstrate the first observation of spin-selective, near background-free and transform-limited photon emission from a resonantly driven QD transition. The hallmark of resonance fluorescence, i.e. the Mollow triplet in the scattered photon spectrum when an optical transition is driven resonantly, is presented as a natural way to spectrally isolate the photons of interest from the original driving field. We go on to demonstrate that the relative frequencies of the two spin-tagged photon states are tuned independent of an applied magnetic field via the spin-selective dynamic Stark effect induced by the very same driving laser. This demonstration enables the realization of challenging QIS proposals such as heralded single photon generation for linear optics quantum computing, spin-photon entanglement, and dipolar interaction mediated quantum logic gates. From a spectroscopy perspective, the spin-selective dynamic Stark effect tunes the QD**

**spin-state splitting in the ground and excited states independently, thus enabling previously inaccessible regimes for controlled probing of mesoscopic spin systems**.

Controlled exchange interaction between confined electron spins in electrically defined semiconductor QDs [1] has proven its worth for QIS [2] within the last decade through numerous key demonstrations - most notably the coherent manipulation of one- and two-electron spins [3–5], spin readout [6], and control of mesoscopic spin reservoirs [7, 8]. Confined spins in self-assembled semiconductor QDs, relying on material-based confinement, are equally promising to serve both as stationary qubits for QIS [9–12] and as probes for studying mesoscopic physics in the solid state [13]. Additionally, in contrast to their electrically defined counterpart, the self-assembled QD excitations (excitons) are at an energy scale that allows interactions with near-infrared photons. Consequently, they have found a unique seat among the potential candidates for solid-state systems through their ability to accommodate both stationary and flying qubits, as well as to serve as an interface between these two qubit classes [14]. In the realm of flying qubits, milestone achievements to-date include single photon antibunching [15, 16], entangled photon pair generation [17], and cavity quantum electrodynamics in the weak [18, 19] and strong [20–22] coupling regimes. A common feature in all the previous studies is the incoherent pumping of the QD transitions through optical exciton generation in either the host matrix such as GaAs or the quasi-continuum states above the higher lying confined states of the QD. This excitation method is convenient as it spectrally eliminates the intense excitation laser background from the emitted photons from the QD, but it also results in an undesired effect. Specifically, the uncontrolled time jitter on the photon emission times due to the nonradiative relaxation of excitons down to the lowest QD state diminishes the coherence of the emitted photons, thus limits their use in linear optics quantum computing algorithms [23] even with cavity coupling [24]. In an attempt to address this shortcoming, increasing attention has turned to the resonant optical excitation of QDs. The significant challenge here lies in being able to distinguish the resonantly scattered photons from the strong excitation laser field; a task complicated by the laser scattering from the high refractive index host matrix (GaAs) in which the QDs are embedded. While recent second-order photon correlation ($g^{(2)}$) measurements have indeed revealed the existence of resonantly scattered light field from a QD in a cavity [25, 26, 27], access to the actual emitted photons themselves without the excitation laser background has proven elusive.

Transition-selective (i.e. resonant) excitation is important from the QIS prospective since it gives direct access to the spin states of a trapped single electron [28]. To-date optical addressing of single spins has relied mainly on the remarkably powerful differential transmission (DT) technique [29]. These measurements or its reflection-based derivatives [30] have already revealed signatures of dressed state formation under strong laser light excitation [31, 32]. This technique, however, relies on the interference of the coherently scattered dipole field and the resonant excitation laser and ultimately prevents access to the emitted photons. In contrast, resonance fluorescence provides direct measurement of the resonantly scattered photons. It was first observed on atomic sodium [33] in 1975 and more recently on a single molecule [34], but with little prospect of linkage to any spin states. It is clear that an outstanding milestone towards the much-sought after spin-photon link of QIS is the observation of spin-selective resonance fluorescence, i.e. access to the resonantly scattered photons that are correlated with the electron spin state of the QD.

**Resonance fluorescence and the Mollow triplet**

A single InAs/GaAs QD embedded in a Schottky diode heterostructure serves as a charge trap where deterministic loading of individual electrons [35] (or holes [36]) from a nearby reservoir is controlled by an applied gate bias (See Supplementary Information Fig. S1). This study is conducted in the single excess electron regime, i.e. at the centre of the single electron charging plateau. Thus, the two spin states of the electron form the ground state manifold which can be coupled via a resonant optical field to the excited state manifold comprising two electrons (forming a spin singlet) and a hole; typically referred to as the $X^{-1}$ transitions. The excited state lifetime leads to a spontaneous emission rate of $\gamma_{sp}$, as indicated in Fig. 1a. Due to their particular spin configurations both state manifolds are doubly degenerate and a finite magnetic field is generally used to lift this degeneracy via the Zeeman Effect [37]. A quantum system which is prepared in the excited state, relaxes to the ground state by emitting a single photon that is excited-state lifetime broadened and is centred on the bare transition frequency, as displayed by the single Lorentzian spectrum of Fig. 1a. In stark contrast, the coupling of the two state manifolds by a resonant monochromatic laser beam results in scattered photons with distinct spectral and coherence properties that may be tuned as the properties of the laser (e.g. power, frequency, and polarization) are varied. Barring any dephasing mechanisms, at low laser powers where the effective Rabi frequency is much less than the spontaneous emission rate, the interference between the scattered photons and the laser field forms the basis of the above-mentioned DT technique. As the excitation power is increased, so that the effective Rabi frequency is stronger than the spontaneous emission rate, the laser dresses the bare electronic states [38] resulting in a composite quantum mechanical system with a modified energy structure. Figure 1b illustrates the formation of these hybrid quantum states from the two original bare states with a red-detuned strong laser field. The original single bare transition energy evolves under the influence of the laser power to four possible transitions in which two are degenerate. The spectrum of the photons scattered, illustrated in the multi-Lorentzian spectrum of Figure 1b, is the well known Mollow triplet [39]. The central frequencies for these three transitions are given by:

$$\nu_{red} = \nu_0 + \Delta - \Omega; \quad \nu_{central} = \nu_0 + \Delta; \quad \nu_{blue} = \nu_0 + \Delta + \Omega; \tag{1}$$

and the linewidth of each detuned sideband is given by $(\tfrac{3}{2})\gamma_{sp}$ in the strong excitation laser regime. Here $\nu_0$ is the bare QD X$^{-1}$ transition frequency (the zero-frequency detuning point in Fig. 1b), $\Delta$ is the frequency detuning of the driving laser from $\nu_0$ (defined as negative for red detuning) and $\Omega = \sqrt{\Delta^2 + \Omega_b^2}$ is the effective Rabi frequency with $\Omega_b$ indicating the bare Rabi frequency.

In order to observe such spectral structure in photons resonantly scattered from a single InAs QD a moderate-finesse (~750) two-mirror Fabry-Perot cavity with a 34.5 MHz spectral window of transmission is used to record the scattered light intensity as a function of cavity transmission frequency (see Figs. S2 and S3 in Supplementary Information). Figure 2a presents X$^{-1}$ resonance fluorescence spectra on a linear-log scale for a range of driving laser power. The laser is set to be resonant with the doubly degenerate bare X$^{-1}$ QD transition frequency, which was determined from our DT measurements. For laser powers above 216 nW two equal weight sidebands emerge, which, together with the central feature, constitute the Mollow triplet. The two sidebands can be understood as arising from radiative transitions between the outer (blue arrow) and inner (red arrow) rungs of the Jaynes-Cummings ladder in Fig. 1b. The strong central feature is the result of the unsuppressed residual laser leaking through our cross-polarizer instrumentation (see Fig. S2 in Supplementary Information). Peak-to-peak sideband splittings determined from this data set are plotted in Fig. 2b against the square-root of the laser power. In the strong excitation regime of $\Omega_b \geq \gamma_{sp}$, the linear fit indeed confirms the expected dependence of the sideband splitting on the effective Rabi frequency, i.e. the square-root of the laser power [39].

Figure 2c is a linear-linear scale zoom-in to highlight the two Mollow sidebands when X$^{-1}$ is driven by a 1.852 μW laser. When fitted with the expected multi-Lorentzian resonance fluorescence spectrum (see Supplementary Discussion), a transition linewidth of 343 (±39) MHz per sideband can be extracted, in close agreement with pure spontaneous emission rate of 356 (±11) MHz obtained from photon-correlation measurements. In the presence of additional fast dephasing Markovian mechanisms, the transition linewidth is no longer determined by the pure spontaneous emission rate and is broadened to $\gamma_{total} = \gamma_{sp} + 2\gamma_{dephasing}$ [40]. The fitted sideband linewidth suggests that such dephasing mechanisms, if they at all exist in this regime of operation, are bounded by a 18-MHz maximum possible rate, set by our experimental uncertainty. This finding alone suggests strongly that emission from the triplet sidebands of the optically dressed QD transition is a predominantly radiation-broadened process yielding *near* transform-limited photons; a pre-requisite for all proposed linear optics QIS applications. We plan both first- and second-order coherence experiments for a more precise measurement of the deviation from the Fourier-transform limit for the sideband photons.

Collection efficiency is as important a factor for photonic QIS applications as spectral purity, therefore an estimate of the photon number collected from sideband emission is necessary. Integrating across the blue sideband and book-keeping for the cavity transmission and detector efficiencies, an estimated 45,000 photons per second per sideband reach the input of our two-mirror cavity. At this particular excitation power, 86% of the emission is coming from the sideband and about 14% is coming from the tail of the central Lorentzian of resonance fluorescence. However, control measurements performed at a gate bias where the excess electron is unloaded back to the reservoir results in the closest possible QD transition being detuned by around 1290 GHz from the laser frequency. The cavity is scanned across the laser in the absence of any $X^{-1}$ transitions reveal that an impressive 99.6% of the total photons is coming from the QD emission directly. For all laser powers above 617 nW, when the two sidebands are clearly separable, the total number of photons collected per sideband remains constant confirming the absence of significant laser background in the above-quoted photon number. Straightforward technical improvements in photon collection efficiency [41] and suppression of amplified spontaneous emission of the excitation laser will further better this figure of merit.

**Frequency control of resonance fluorescence spectrum**

In addition to the laser power, the sideband spectrum may be tuned by controlling another externally accessible degree of freedom; the laser frequency. Figure 3a is the plot of Eq. (1) for both sidebands as a function of the laser frequency detuning ($\Delta$) from the bare $X^{-1}$ resonance at fixed laser power. In Fig. 3b the measured $X^{-1}$ resonance fluorescence spectra, driven by 1.852 μW laser, are plotted as the laser frequency is tuned across the bare QD transition frequency. Measuring the relative spectral separation of the red sideband, when the laser is red detuned by 2.48GHz, from the blue sideband, when the laser is blue-detuned by 3.32 GHz, it is possible to achieve resonantly driven photon emission across a frequency band of ~ 14 GHz. This is an impressive range that is 40 times larger than the 356 MHz spontaneous emission rate, and is by no means an upper limit, but can be further increased by laser power and detuning. In order the relate this range to other tuning mechanisms, we emphasize that 16 GHz is the range obtainable via DC stark shift of the $X^{-1}$ transition throughout the whole single electron charging plateau [11]. Alternatively, this is the same frequency shift that each of the two degenerate $X^{-1}$ transitions experiences under an applied magnetic field of 1 Tesla [37, 42]. The sideband splittings extracted from the data set of Fig. 3b, is plotted in Fig. 3c as a function of laser detuning. The red fit curve with the functional form of $2\sqrt{\Delta^2 + \Omega_b^2}$ is used to determine a bare Rabi frequency of 2.76±0.2 GHz. From this Rabi frequency value we determine a dipole moment of 27.8±0.2 Debye in agreement with the value extracted from our DT measurements.

Although the most distinct signature of resonance fluorescence is the three spectrally isolated Lorentzian lineshapes, this spectrum occurs only when the bare Rabi frequency is considerably larger than the spontaneous emission rate. If the relationship between the two rates is reversed, the spectrum collapses to a single squared Lorentzian. However, another distinct spectrum may be realized when the excitation laser has the same low power, but sits at a finite detuning from the bare transition frequency. In this case, the fluorescence spectrum exhibits neither single nor triple Lorentzian lineshape, but rather has a double peaked structure, as predicted by Mollow [39], and is explicitly given by the second term of

$$F(\nu) \propto \delta(\nu - \nu_{\text{laser}}) + 1/\left[(\nu - \nu_{\text{laser}} - \Delta)^2 + \gamma_{\text{sp}}^2/4\right] \times \left[(\nu - \nu_{\text{laser}} + \Delta)^2 + \gamma_{\text{sp}}^2/4\right]. \tag{2}$$

The first term in Eq. (2) is the spectral component that has full first-order coherence and dominates the spectrum at low excitation regime. This coherent component follows the excitation laser frequency and can interfere with the laser (this is the basis for DT). The second term is the incoherent component, i.e. resonance fluorescence. Interestingly, despite being far from the strong excitation limit, the resonance fluorescence spectrum is still determined by the excitation laser frequency.

In order to explore this low-power limit, the fluorescence spectrum originating from a 230-MHz blue-detuned excitation laser fixed at a power of 0.512 nW is presented in Fig. 3d. The corresponding Rabi frequency for this laser power is ~50MHz which is roughly 7 times smaller than the spontaneous emission rate measured in $g^{(2)}$. To verify the double peaked structure of Eq. (2), the measured spectrum was fit by fixing both the detuning to 230 MHz and the spontaneous emission rate to 343 MHz. The result is the purple solid curve displayed in the plot labelled "1". For comparison, in the plot labelled "2" the same data set is fit using the three-Lorentzian resonance fluorescence spectrum fixing the Rabi frequency, detuning and spontaneous emission rate to their known values. The mismatch in plot labelled "2" supports our observation of the spectrum given by Eq. (2).

We would like to point out another interesting regime obtained via detuning. When the excitation laser frequency ($\nu_{laser}$) is considerably far red-detuned, the red (blue) sideband frequency approaches $2\nu_{laser} - \nu_0$ ($\nu_0$). This is indeed the analogous regime studied by Aspect *et al.*, on single atomic transitions [43]. In this seminal work, photon correlation measurements between these two sidebands revealed that the emission dynamics of the two sidebands in this regime is substantially different from that obtained in the near-resonant laser excitation. Specifically, the red-sideband photon has to be generated from two virtual states, while the blue-sideband photon is emitted from the two real atomic states. Due to the complex nature of this scattering process, the first photon to be emitted is hence always in the blue sideband, followed instantaneously by a photon in the red sideband. This regime of operation has immediate impact on *heralded* (on-demand) generation of transform-limited photons from the red-detuned sideband, when triggered by the detection of a photon from the blue-detuned sideband.

**Mollow quintuplet and the spin-selective dynamic Stark effect**

So far all measurements were performed in the absence of a magnetic field. We now apply a 50 mT external magnetic field along the sample crystal axis in order to break the X$^{-1}$ spin degeneracy. Independent DT measurements under this magnetic field reveal a 1-GHz Zeeman splitting between the two orthogonally polarized (in the circular polarization basis) transitions. As simulated in Fig. 4a, the two dressed Zeeman split sidebands (the blue and red solid lines) are directly correlated to the spin state of the electron and are further tuned differently by the excitation laser frequency. In what follows all frequencies are referenced to the zero magnetic field bare QD X$^{-1}$ resonance. First, the linearly polarized pump laser is fixed to 1.852 µW power and its frequency is red-detuned by 1.25 GHz (line cut 3 in Fig. 4a). In Fig. 4b the measured resonance fluorescence spectrum exhibits a distinctive five-Lorentzian structure beyond the previously discussed triplet and each sideband transition is *directly linked with a QD* *electronic spin state*. In total there are 6 features in the spectrum, but, much like the Mollow triplet, the central line is comprised of two degenerate transitions locked to the detuned laser frequency, thus we observe a spectral signature for the Mollow quintuplet. In Fig. 4b, in addition to the spectral location of each sideband, we also indicate the electron spin associated with the transition. In Fig. 4c, the spectrum of light scattered from the blue-detuned sideband (the blue box in Fig. 4a) is plotted as the laser frequency is varied under a fixed magnetic field of 50 mT. The five spectra presented here arise when the laser is red-detuned by 1.75, 1.5, 1.25, 0.75, and 0 GHz, from left to right, respectively. By varying the laser detuning, at constant power, the Zeeman splitting of the transitions induced by the magnetic field can be altered (panels 1-4 of Fig. 4c) and ultimately even cancelled (panel 5 of Fig. 4c).

What we demonstrate here is a combined outcome of the Zeeman and dynamic Stark effect [32, 44, 45], which allows us to tune independently the energy splitting of the ground and excited states. For InGaAs QDs, the electron and hole g-factors are known to be around -0.6 and 1.4, respectively [37]. Therefore, the ground and excited state manifolds respond differently to the applied magnetic field. The dynamic Stark effect, however, is independent of either manifold's Landé factor. Consequently, the two state manifolds in this regime display level splittings corresponding to an effective Landé factor tuned by the properties of the excitation laser. This optical controllability of the spin ground states could alternatively be coined the optical Zeeman effect. The essence of this effect lies in the imbalance of the effective Rabi frequencies ($\Omega = \sqrt{\Delta^2 + \Omega_b^2}$) experienced by the two spin transitions. In this work we utilize the $\Delta$-dependence to enforce this imbalance and show the cancellation of the magnetic field induced spin-splitting. An alternative approach is also possible through the excitation laser polarization allowing a $\Omega_b$ imbalance in conjunction with the optical selection rules of the spin transitions. The condition given in panel 5 of Fig. 4c is particularly interesting, since both the ground and the excited states are identically split resulting in an effective Landé factor of 0.4. The consequence is the generation of spectrally indistinguishable photons with well-defined spin tags in their circular polarization state for *any* applied magnetic field strength. This is a regime that merits a thorough investigation for the challenges of spin-photon entanglement [46, 47].

**Conclusions and Outlook**

In this work, we present the first ever observation of both radiative-lifetime broadened and background-free resonance fluorescence from a single QD transition. We demonstrate that the frequency of the emitted photons can be tuned by presenting an example of 14 GHz of shift via fixing the excitation laser power and detuning its frequency. Finally, resonance fluorescence is combined with the dynamic Stark effect to link the spectrum of the emitted photons to the spin state of the electron. We show under a modest magnetic field of 50 mT we are able to control the spectral separation of the spin-tagged photons anywhere from the full expected Zeeman splitting to zero separation yielding full spectral overlap and even to reverse ordering. This ability may have direct implications to feasible realization of spin-photon entanglement schemes as well as independent control over ground and excited state energy splitting, i.e. Landé factor control under a given magnetic field breaking the restrictions brought on by the material-manifested Landé factors for each charge in these systems.

Propelled by the narrow linewidths observed here our immediate research directions include the measurement of both first- and second-order interference properties of each sideband independently to confirm the generation of transform-limited photonic qubits without requiring any cavity coupling [48]. In parallel, the optical tuneability and the observation of emission correlations between the two sideband photons (showing the time ordering of the emission due to the Jaynes-Cummings ladder cascade) may serve both as a heralded photon source and as a sensitive spectroscopic probe for weakly coupled states near the original QD transitions.

This step may prove more useful also when used in conjunction with tunnel coupled QD pairs, thus probing more complex ground and excited state manifolds. Finally, the modest magnetic field value used in these experiments is chosen in order to avoid any significant spin pumping and nuclear spin polarization effects, while spectrally resolving the two spin-selective transitions. Therefore, how the emission spectrum of the dressed states evolves when one or both of these mechanisms are significant is an interesting topic of research on its own. Each one of these next steps, if achieved, is expected to have direct impact on both the pursuit of solid-state QIS and our understanding and control of mesoscopic physical systems.

**Methods**

The InAs/GaAs quantum dots studied were grown by molecular beam epitaxy (MBE) using the partially covered island technique and are embedded in a Schottky diode heterostructure. An illustration of the sample structure is in Fig. S1. A diagram of the experimental setup can be found in Fig. S2. For the measurements the gated QD sample is housed in a magneto-optical bath cryostat and cooled to 4.2 K. A cubic zirconia solid immersion lens (SIL) is mounted on the epitaxial sample surface in order to improve both the light focusing and light gathering power of the fiber-based confocal microscope. We first identify an $X^{-1}$ transition from voltage dependent photoluminescence with an excitation wavelength of 780 nm (PL setup in Fig. S2). Then the differential transmission laser spectroscopy technique is used to determine the exact spectral location of the $X^{-1}$ transition (DT setup in Fig. S2). Next, a scanable single mode diode laser with 1.2-MHz frequency and 0.5% power stabilization tuned to the $X^{-1}$ transition. The light resonantly scattered by the QD is collected through the second arm of fiber confocal microscope, sent through a ~34.5 MHz frequency stabilized Fabry-Perot cavity and subsequently analyzed with a liquid nitrogen cooled CCD (RF setup in Fig. S2). The Fabry-Perot normalized transmission is presented in

Fig. S3. The measured Fabry-Perot throughput is 25%. To suppress the background laser light and collect the resonance fluorescence spectrum we operate the microscope in a dark-field configuration by placing a polarizer perpendicular to the incoming linearly polarized laser field. In this configuration we measure an extinction of the laser light equal to $5 \times 10^3$. In order to obtain the excited state lifetime, Hanbury-Brown and Twiss type photon-correlation measurements are performed by two single photon counting avalanche photodiodes and a record of coincidence events is kept to build up a time-delay histogram ($g^{(2)}$ setup in Fig. S2).

**Acknowledgements** We thank A. Imamoğlu for his continuing support and guidance, S. Faelt and A. Badolato for providing high-quality QD samples, G. Burkard, J. M. Taylor, H. Türeci, A. K. Swan, M. S. Ünlü, B. B. Goldberg and J.-W. Pan for insightful discussions. This work was supported by grants and funds from the University of Cambridge and EPSRC Science and Innovation Awards. A.N.V. is supported by QIPIRC of EPSRC, Y.Z. is supported by A. v. Humboldt Foundation and LGFG, and C.-Y.L. is supported by Univ. of Cambridge, CSC and CAS. Correspondence and requests for materials should be addressed to M.A. (ma424@cam.ac.uk) and A. N. V. (anv21@cam.ac.uk).

**Figure Captions**

**Figure 1 | Energy diagram for the QD $X^{-1}$ transitions. a,** The doubly degenerate 2-level system comprised of a single electron ground state and the trion (electron spin singlet and a single hole) excited state in zero magnetic field. The plot presents the simulated spectrum of spontaneous emission from a 2-level system. **b,** Jaynes-Cummings ladder structure for the dressed two-level system in a strong laser field which is red-detuned by $\Delta$ from the resonance. Four possible transitions are indicated, two of them frequency degenerate. The resulting fluorescence spectrum exhibits the Mollow triplet as illustrated in the corresponding plot for zero-detuning condition.

**Figure 2 | Power dependent resonance fluorescence. a,** Evolution of the Mollow triplet spectrum as the resonant laser power is increased from 0.512nW to 1.852μW. The intensity of the spectrum is plotted in logarithmic scale. Each data point is recorded for 60 seconds. **b,** Extracted sideband splitting as a function of pump field strength, i.e. the square root of the pump power, on a linear scale. The data points follow the expected linear relationship between square root of laser power and the Rabi frequency determined sideband splitting confirming the dressed-state picture of Fig. 1b. **c,** Zoom-in plot of the 1.852 μW fluorescence spectrum sidebands with a linear intensity scale. The boxes highlight the sidebands from which we extract a transition linewidth of 343 (±39) MHz (See Supplementary Information) and a collected photon rate of ~45,000 per second per sideband.

**Figure 3 | Dependence of resonance fluorescence on laser detuning. a,** Simulation of the scattered photon frequencies (red and blue solid curves) for the dressed states of an $X^{-1}$ transition as a function of laser detuning.  The dashed green line corresponds to the

bare X$^{-1}$ transition and the black solid line indicates both the laser frequency and the central peak of fluorescence. **b,** Experimentally observed resonance fluorescence spectrum as a function of laser frequency detuning from the bare X$^{-1}$ resonance. The laser power is fixed at 1.852 µW for all detunings. The central peak, which overlaps with the laser, is removed for clearer presentation of the sideband behaviour. The red and blue dashed curves tracing the sidebands are guides to the eye commensurate with the red and blue curves in panel a. **c,** Sideband splitting as a function of laser detuning. The bare Rabi frequency of 2.76 GHz and the related transition dipole moment of 27.8±0.2 D are extracted from the fitted red curve. **d,** Linear-scale zoom-in plot of fluorescence spectrum (purple squares) obtained from a 0.512-nW laser with a deliberate 230-MHz blue detuning from the bare X$^{-1}$ resonance. The blue diamonds are the control data set representing the spectrum when the QD electron is unloaded back to the reservoir eliminating all relevant optical resonances. The purple solid curve in top plot labelled "1" is the theoretical fit to the purple squares including a single Lorenztian with 34.5 MHz linewidth for the unsuppressed laser and the coherently scattered component (dashed curve), as well as the additional two-peaked spectrum given in Eq. (2) for the resonance fluorescence component (red curve) with detuning and spontaneous emission rate fixed to 230 MHz and 343 MHz, respectively. Plot labelled "2" is the same data set but fitted with the conventional three-Lorentzian spectrum for resonance fluorescence (purple curve). The three Lorentzian components are displayed independently as red, blue and black curves. The strong mismatch with the well-known Mollow triplet is, therefore, clear in this regime.

**Figure 4 | Mollow quintuplet and the spin-selective dynamic Stark effect. a,** The simulated scattered photon frequencies for the 1-GHz Zeeman split X$^{-1}$ transitions under a 50-mT external magnetic field. The blue solid lines are the dressed sidebands

corresponding to the blue shifted bare Zeeman transition (dashed blue line) and the red solid lines are the dressed sidebands corresponding to the red shifted bare Zeeman transition (dashed red line). The ends of each line are decorated with an illustration indicating the specific QD spin ground state for each transition. The blue box and the vertical solid black lines highlight the spectral window and laser detunings we experimentally investigate in panel c. **b,** The resonance fluorescence quintuplet under application of a 50 mT magnetic field and a 1.25 GHz red detuned laser with a power of 1.852 µW. Each sideband splits into a doublet in which each transition is linked to a specific QD spin state, where the blue (red) peak indicates the spin-up (down) state. **c,** The evolution of the blue-detuned Mollow sideband spectrum for a series of laser frequency detunings. The inset in the upper left corner illustrates how the laser detunes from the blue (red) Zeeman split transition. The number in the upper right corner designates the corresponding line cut indicated in panel a. The external magnetic field for all spectra is fixed to 50 mT, and the change in spin-splitting originates from laser detuning only.

**Figure 1:**

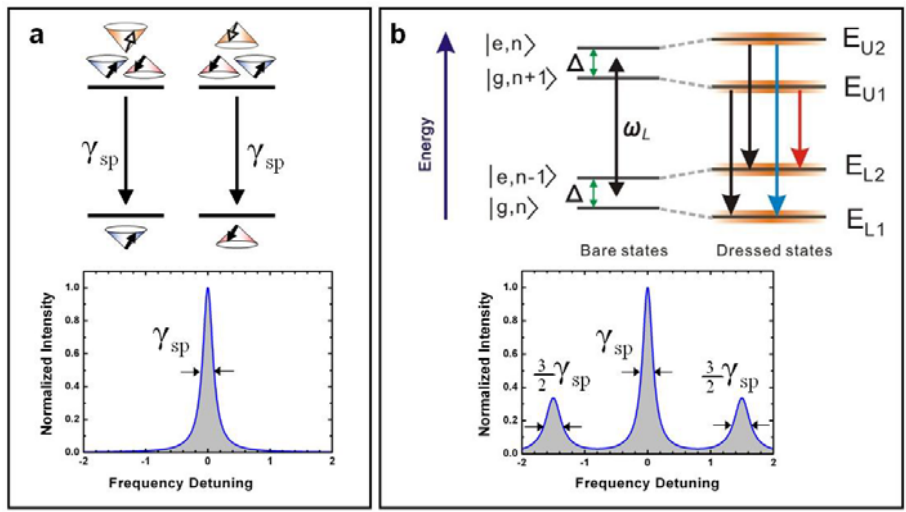

**Figure 2:**

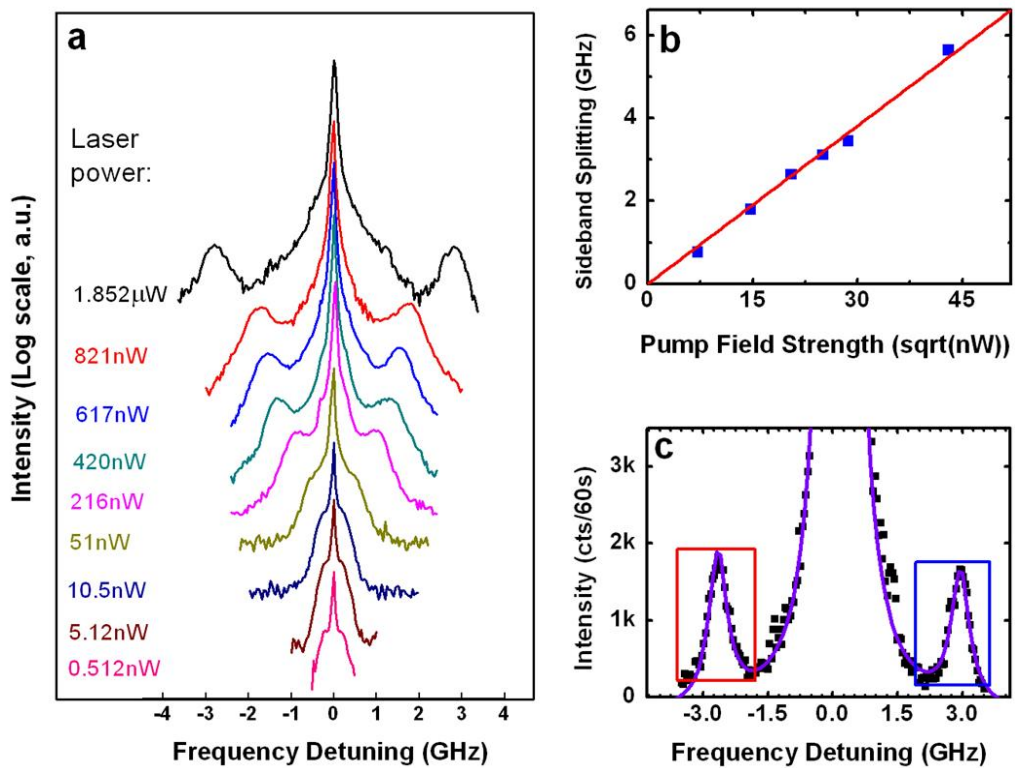

**Figure 3:**

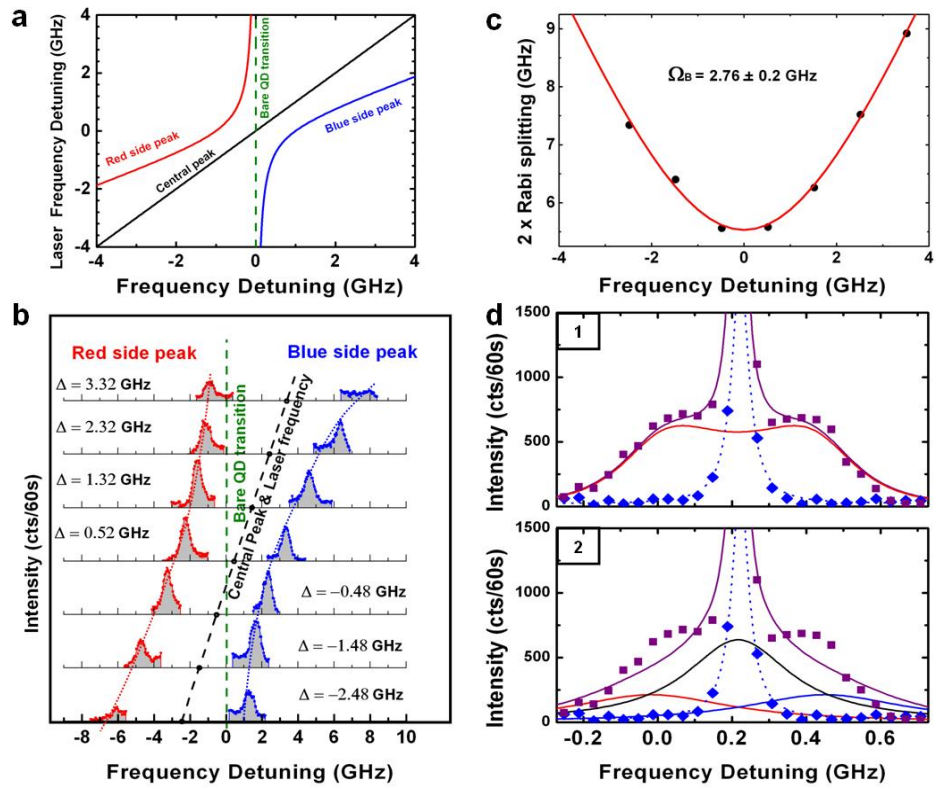

**Figure 4:**

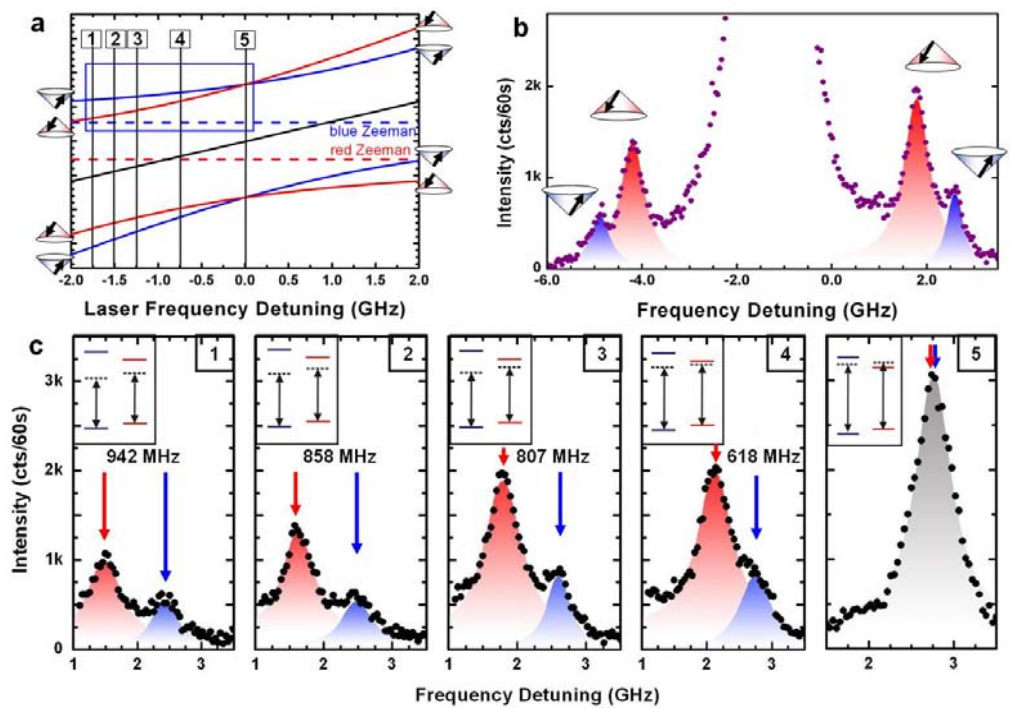

## Supplementary Information:

This document contains supplementary figures providing information on the QD sample structure (Fig. S1), key components of the experimental setup (Fig. S2), experimental verification of the spectral transmission window for the Fabry-Perot cavity used in these experiments (Fig. S3), and a discussion section on optical transition linewidths measured using different techniques (Fig. S4).

### A Discussion on Transition Linewidth:

This is a short supplementary note on the transition linewidth measured using conventional nonresonant photoluminescence (PL), differential transmission (DT), resonance fluorescence (RF) and photon correlation (g$^{(2)}$) techniques. In all linewidth discussions in this manuscript we use the below definitions. All spectral measurements result in Lorentzian lineshapes and the intensity correlation measurement results in a double exponential profile.

The functional form of a Lorentzian is:

$$\propto \frac{1}{\Delta^2 + (\gamma/2)^2}$$

where $\Delta$ is the independent variable such as laser-transition detuning in DT or emission frequency in PL and RF, and $\gamma$ determines the transition linewidth. The measured full width at half maximum (FWHM) of this Lorentzian is $\gamma$. For optical transitions in which the only dissipation and/or dephasing mechanism is spontaneous emission, which we assume throughout this discussion, $\gamma$ is equal to the spontaneous emission rate $\gamma_{sp}$ [1, 2]. In PL the emission line shape $L$ is [1]

$$L(\omega_{em}) \propto \frac{1}{\Delta^2 + (\gamma_{sp}/2)^2}$$

where $\Delta$ is the detuning between the emitted light frequency $\omega_{em}$ and the transition resonance frequency. We highlight in PL the emission linewidth is exactly equal to $\gamma_{sp}$.

**1) DT:** In DT the extinction line shape is measured directly and should equal [3]

$$\Delta T(\omega_{sc}) \propto 1 - \frac{1}{\Delta^2 + (\gamma_{sp}/2)^2}.$$

The important point here is that the linewidth is now equal to $\gamma_{sp}$.

**2) RF sidebands:** In RF, in the strong excitation laser limit, the side band lineshape is [2]

$$L(\omega_{em}) \propto \frac{1}{(\Delta - \Omega)^2 + (3\gamma_{sp}/4)^2}$$

where $\Omega$ is the bare Rabi frequency. In RF the sideband linewidth is equal to $3\gamma_{sp}/2$.

**3) RF central peak:** Finally, in RF the centerband lineshape, assuming strong excitation, is [2]

$$L(\omega_{em}) \propto \frac{1}{\Delta^2 + (\gamma_{sp}/2)^2} \ .$$

The linewidth, as in DT, is equal to $\gamma_{sp}$. Due to the laser background this peak is inaccessible in the strong excitation regime.

**4) $g^{(2)}$ profile:** In photon correlation measurements, the correlation lineshape for a single anharmonic optical transition is [2]

$$g^{(2)}(\tau) \propto 1 - \exp(-\gamma_{sp}\tau) \ .$$

This is the well-known anti-bunching dip in the limit of vanishing excitation laser power to eliminate multiple carrier relaxation effects. A fit to the $g^{(2)}$ anti-bunching dip reveals $\gamma_{sp}$.

**Supplementary Figure Captions**

**Figure S1 |** An illustration of the InAs/GaAs sample structure. All experiments discussed here were performed on single QDs isolated in real space and spectrum on the top QD layer (red), with no optical signatures of any other QDs from the lower QD layer (blue).

**Figure S2 |** A schematic of the experimental setup indicating key elements comprising each technique used in the manuscript.

**Figure S3 |** The measured transmission linewidth of the Fabry-Perot cavity as it is scanned across a 300-Khz linewidth single frequency laser independently stabilized to 1 MHz frequency uncertainty.

**Figure S4 |** A comparison of the linewidths extracted from **a,** power broadening via Differential Transmission **b,** photon correlation [g$^{(2)}$] at low power incoherent excitation **c,** nonresonant Photoluminescence transmitted through the same Fabry-Perot cavity used in RF measurements and **d,** the resonance fluorescence sideband at 1.852 μW excitation laser power.

**Supplementary Figure 1:**

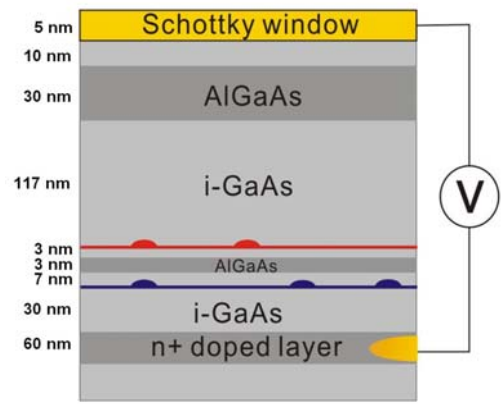

**Supplementary Figure 2:**

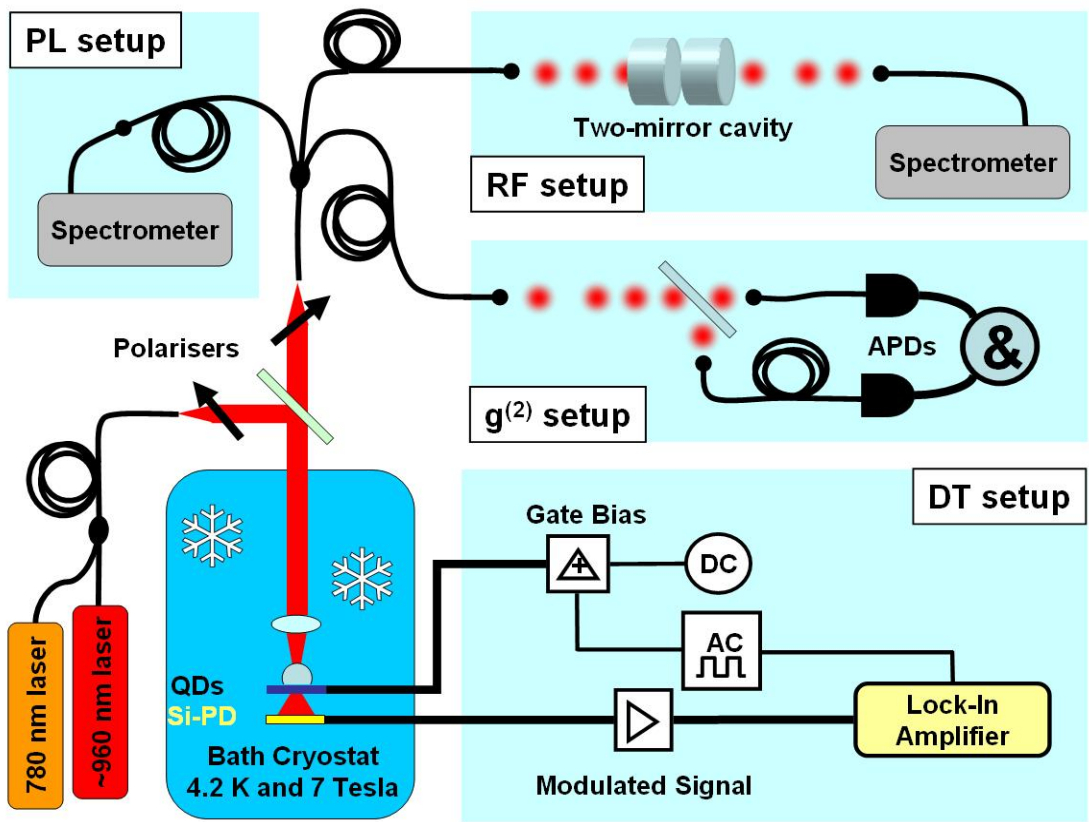

**Supplementary Figure 3:**

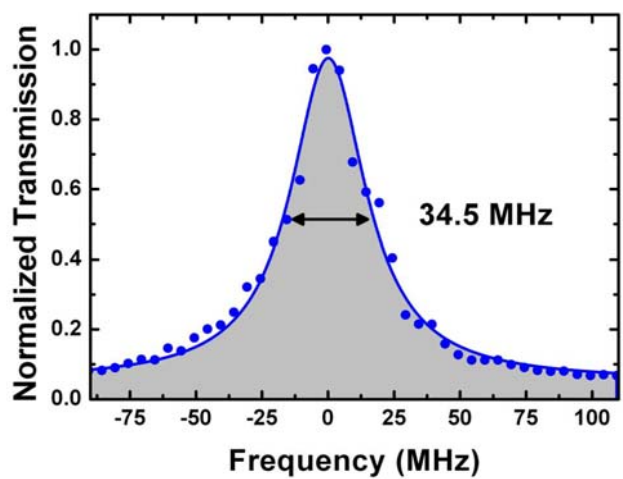

**Supplementary Figure 4:**

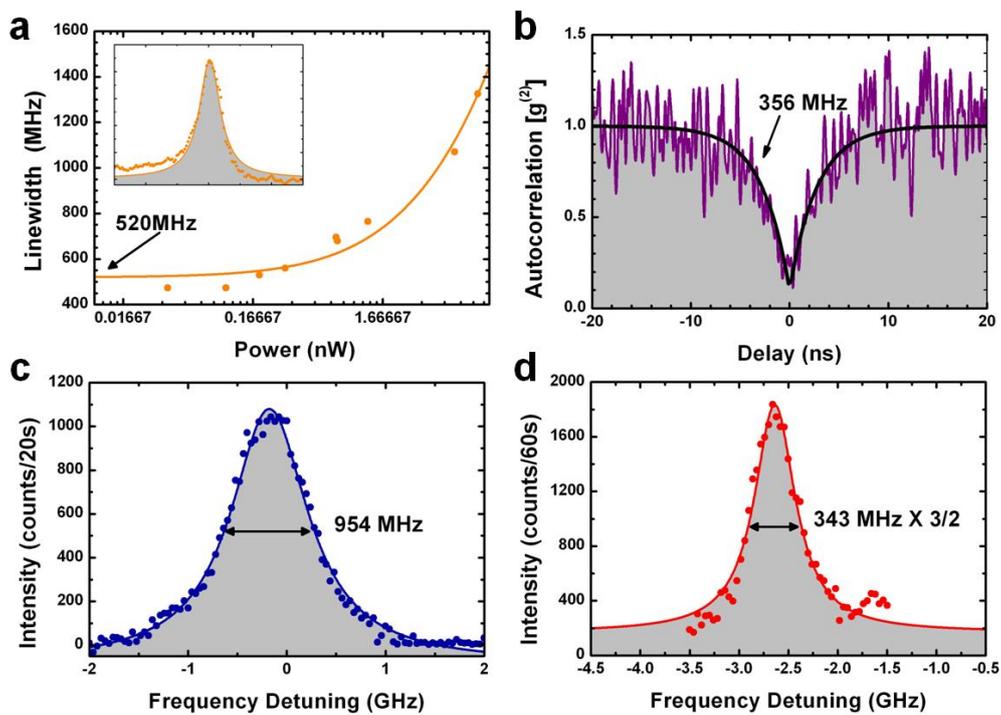